"*Indigenous Astronomy – Best Practices and Protocols for Including Indigenous Astronomy in the Planetarium Setting*"

Annette S. Lee, Nancy Maryboy, David Begay, Wilfred Buck, Yasmin Catricheo, Duane Hamacher, Jarita Holbrook, Ka'iu Kimura, Carola Knockwood, Te Kahuratai Painting, Milagros Varguez

Annette S. Lee aslee@stcloudstate.edu
Nancy C. Maryboy dilyehe@gmail.com
David Begay dbegay@gmail.com
Wilfred Buck wilfredb@mfnerc.com
Yasmin Catricheo YCatricheo@aui.edu
Duane Hamacher duane.hamacher@gmail.com
Jarita Holbrook astroholbrook@gmail.com
Ka'iu Kimura lesliek@hawaii.edu lesliek@hawaii.edu
Carola Knockwood Carola.Knockwood@novascotia.ca
Te Kahuratai Painting tekahuratai@gmail.com
Milagros Varguez milagrosvarguez@gmail.com

**Abstract**
Indigenous people have nurtured critical relationships with the stars, from keen observation and sustainable engineering to place-based ceremony, navigation, and celestial architecture for tens of thousands of years. The Indigenous relationship and knowledge of the sky is exceptional in that it encompasses mind, body, heart, and spirit.

This panel is organized by the IPS's newly formed Indigenous Astronomy Working Group. It brings together Planetarium Professionals, Indigenous Star Knowledge Keepers, Indigenous Astronomy experts, Cultural Astronomers, and allies of Indigenous STEM communities from Canada, U.S., and internationally to discuss best practices for dissemination of indigenous astronomy specifically for science communicators and the planetarium community.

**I      Introduction**

Astronomy as presented in planetarium shows can be beautiful but ultimately distant with no way for audiences to link what they are viewing to their everyday lives. In contrast, Indigenous astronomy has people at its center. It is about people, relationships and the sky, not just about the sky. As eloquently described by two well-known Dine scholars:

> *The Dine word, 'Sitsoii Yoo' (Star or original light that evolves) acknowledges the ancient relationship to the original light that came from the original star. It acknowledges all life, including human life, proceeded by the original energy of light, similar to astrophysicists' explanation that we are stardust. Ancient teachings tell us that when humans look at the Milky Way galaxy at night, they are actually looking at themselves, from which energies they actually evolved. (Maryboy and Begay 2020)*

Clearly there is an enormous amount of wisdom in our Indigenous Knowledges Systems (IKS) and in particular, our Indigenous relationship with the night sky. Indigenous peoples around the world, from time immemorial, have seen the heavens, obtaining from their perspectives information that has helped them define areas of each culture, whether from a religious, biological, spiritual or even temporal point of view. And yet the enormity of post-colonization can and should not be understated. We might be living five-hundred years (c. 1492) after the Portuguese and Spanish armadas first sailed west and bumped into the North and South American continents, but the impact from the loss of cultural knowledge and language is a present, harsh, and enduring reality.

Fortunately, not all has been lost. Elders like Nancy Maryboy and David Begay began working on revitalization of Dine' Indigenous Astronomy over three decades ago. Other Indigenous voices from North America (such as Wilfred Buck and Annette S. Lee), from the Hawaiian Islands (Ka'iu Kimura and Kalepa Baybayan), from Mexico (Milagros Varguez), have been working together with African American/African Indigenous Astronomy scholars (like Jarita Holbrook), Maori Knowledge Keepers (such as Te Kahuratai Painting), Aboriginal Australian & Torres Strait Island communities (through Duane Hamacher), and more recently, Chilean-Argentine Indigenous Mapuche people (via Yasmin Catricheo). We are at a critical moment. People have an excitement and curiosity about the night sky, a kind of craving for more, but at the same time we are in danger of dark skies going extinct (Bogard 2014; Deudney 2020; Harris 2009). This moment is an opportunity to recognize that science is embedded with culture and history, and it is past time to widen the lens. All people from all over the world, throughout human history have had a keen relationship with nature and have therefore been practicing science/scientific thinking for millennia.

This document serves to support the Indigenous voice as the lead voice in revitalizing Indigenous star knowledge, starting with our own Indigenous communities. We offer this document as an introduction to present best practices and protocols for how mainstream institutions such as (non-Indigenous led) planetariums can support our efforts in an authentic and meaningful way.

## II      What is Indigenous Astronomy?

The first step in creating a more inclusive, diverse, and equitable learning environment in planetariums, and indeed in all of STEM is to acknowledge that science is by definition embedded with culture. Culture is more than superficial or performative markers like holidays, customs, and ethnic food. Culture is often unconscious but has a huge impact on the philosophical underpinnings of a society (Lee 2020). Similarly, science has become the ever narrowing, laser focused, divide-and-conquer endeavor that we know today only in the past few hundred years. For example, the entomology of the word 'physics' comes from 'physica' or 'physicks' which is from Latin meaning 'the philosophy of nature or natural philosophy' (Oxford English Dictionary 2019). Certainly we can recognize that the multitude of peoples that have existed on Earth have had different ways of 'relating to nature'.

Similarly, a person's worldview is learned from the environment in which the person grows. As part of the process of inculturation, the newborn begins to learn not only the language and customs but also the basic assumptions, premises, and concepts of his parents, family, and community (Chate 2017). At the moment of removing a person from his culture (deculturation), they lose their identity and with it the chain of transmission of the knowledge of their ancestors to generation after generation is broken. By acknowledging one and only one culture's natural philosophy we are losing valuable perspectives related to the physical/natural world, acculturation/inculturation, environmental wisdom as well as methods of transmission of knowledge. Chilean researcher Michel Duquesnoy explains that people are intentionally denying their Indigenous origin because they want to appear 'pure' and because Indigenous people are still viewed with mistrust. He explains the importance of the state's responsibility in developing laws that respect Mapuche culture:

> *The problem is due to weak policies and the blindness of a somewhat discriminating society, dialogue is hindered or eliminated, which is essential in the development of a society and in the creation of a common identity.*
> (Duquesnoy n.d.)

Whether we use the more formal term 'Indigenous Knowledge Systems' (IKS), or the more general term 'worldview', we can recognize that Indigenous people had and still have a keen awareness of the natural world. Relationships and participation with the natural world were and still are key elements in Indigenous science.

The term "Indigenous Knowledge Systems (IKS)" is rooted in the voice of post-apartheid South Africa, and similar movements followed in Australia, New Zealand, and more recently Canada. The South African government began to recognize IKS as a critical and valuable perspective worth protecting and promoting at the federal level. Indigenous African scholars like Munyaradzi Mawere, Lesley LeGrange, and Meshach Ogunniyi (Le Grange 2007; Mawere 2015; Ogunniyi 2005) give the following outline of the differences between Indigenous Knowledge System and Western Science:

*Table 1 – Comparison of Indigenous Knowledge Systems with Western Science, Le Grange 2007, 585*

| Indigenous Knowledge Systems: | Western Science: |
|---|---|
| nature is real and partly observable | …versus nature is real and observable |
| events have both natural and unnatural causes | … versus all events have natural causes |
| the universe is partly predictable and partly unpredictable | … versus the universe is predictable |
| language is important as a creative force in both the natural and unnatural worlds | … versus language is not important to the workings of the natural world |
| knowledge is a critical part of culture | … versus science is culture free |
| humans are capable of understanding only part of nature | …versus humans are capable of understanding nature |

Additional elements of Indigenous science are given by North American Blackfoot elder and scholar, LeRoy Little Bear, (1) all things are made of energy, "In Aboriginal philosophy, existence consists of energy. All things are animate, imbued with spirit, and in constant motion" (Little Bear 2000, 77); (2) we are all related, "In this realm of energy and spirit, interrelationships between all entities are of paramount importance…" (Little Bear 2000, 77). To reiterate this critical point, Lakota elder, Albert Whitehat writes,

> *Philosophically, 'mitakuye oyasin' states that a person is related to all Creation. 'Mitakuye oyasin' reminds us that we all come from one source, the blood of Inyan, and therefore we need to respect each other….Through these ways we maintain 'wolakota' peace… During pre-reservation times there was only one philosophy, one culture, and one language. We did not separate prayer from our daily life. Everything was the Lakota way of life, and 'mitakuye oyasin' was practiced in all situations.* (1999, 46)

From a South America Indigenous perspective, a term similar to Indigenous Knowledge Systems (IKS) that is well-used in original cultures is 'worldview'. The worldview consists of the assumptions, premises and ideologies of a sociocultural group that determine how they perceive the world" (Lafuente and Sanchez 2010). For example, from the perspective of the Mapuche people of Chile there is the belief that natural forces govern and regulate the universe. Within that universe, humans are part of this structure which is in harmony with everything and in balance between positive and negative. There is a very strong connection with nature and the environment in which one lives. It is cared for, protected, respected and interpreted for the future good of the community in accordance with the past, and with those who are no longer there physically, but whose energy remains in every corner of nature. For this reason, the connection with nature is so important for the Mapuche community. As explained further, nature is not only a source of food and the home that welcomes us, it is also a connection with and representation of the ancients and those who were before us and who we have inherited the knowledge that allows us to continue to grow.

To the point, the Indigenous worldview is at times in direct contrast with the Western worldview. Western worldviews focus on how humans exercise dominion over nature to use it for personal and economic gain; human reason transcends the natural world and can produce insights independently; nature is completely decipherable to the rational human mind (Knudtson and Suzuki 1992). As (Carter et al. 2003: 6) note, "The emphasis of Western science is on mastering, controlling, and transforming nature and promotes individual success and competition."

The damage of colonization is still being dispensed when we put on our blinders and teach science (or astronomy) as 'the objective truth'. As stated by Ininew elder, Wilfred Buck (2012: 73):

> *We as individuals tend to view our civilization as "the best" and when our teachings, knowledge, and belief systems are ridiculed, marginalized and then utterly dismissed as "quaint", we begin to question our world view. This has happened and is still happening to First Nations people as well as all colonized peoples. Until other worldviews are proposed and considered, there will be a distinct "difference" and "quaintness" about all that is not mainstream. In addition, our children will see these differences and attempt to discard them in order to become more mainstream. These teachings reflect the differences and propose another perspective, broadening and giving voice to them...The implications for the educational systems (public schools and federally funded Band operated schools) in which ~~are~~ our children are indoctrinated, is that it recognizes the "otherness", and becomes a part of our multicultural nation.*

Simply put, if astronomy is defined as 'a natural science that studies celestial objects and phenomena' (Unsöld and Baschek 2001), then we can define Indigenous Astronomy as 'a living relationship with nature focused on the sky and celestial phenomena that is deeply embedded in keen observation and participation, anchored to heritage that goes back tens of thousands of years, and includes recognition that all living things are embedded with spirit and therefore related.'

### III.   Importance of Indigenous astronomy?

One important aspect of Indigenous astronomy is how knowledge connected to the sky or about the sky is culturally encoded for sharing as well as remembering. Stories about the sky often provide a mapping of the sky (Holbrook 2014). From the Ewe people of Togo, Orion is the hunter, same as in the Greek myth, but he is hunting chickens (Spieth 1911: 53). Their constellation Dzeretsia fits the description of Orion, with a tattoo on his belly representing three people seems to be describing the belt of Orion, and his genitals seem to be the sword of Orion. Kolkovino (the chickens) as described is a cluster of stars, assumed to be the Pleiades but from the description it could be the Hyades; nevertheless, they are what Dzeretsia is hunting. Dzeretsia is connected with yam planting. He leaves the night sky at the time the yams are seeding, when he appears near the zenith at sunset yams are planted, when he appears in the west after sunset planting is stopped. There isn't a causal link between Dzeretsia and yams given that he is a hunter, but instead he only serves as an agricultural calendar marker. However, Dzeretsia seems to be given some agency over rainy and dry weather: When it is rainy, they say that Dzeretsia has dipped his foot in the water; and when it is dry, they say Dzeretsia is keeping his foot in the fire (Spieth 1911, 53). This example encodes important agricultural information about when to plant yams, thus the location of constellations in the night sky is entwined with the livelihood and food security of the Ewe.

Another example of place-based astronomical knowledge tied to seasonal and cultural agricultural practices is found in the Dine teachings of when to plant corn seeds. This occurs in the spring in Navajoland, when the Pleiades are visible and then disappear into the western horizon. They are spoken of as little boys who disappear behind some hills. That is a signal to begin planting. When the Pleiades become visible in the east, and the little boys reappear in the night sky, it is a sign to stop planting. Elders say, "Never let Dilyehe (the Pleiades) see you plant your seeds!" (Maryboy and Begay 2005, 2010, 42-3).

Constellations, or Waŋlen in Mapuche culture, were used as important calendar events and for deep spiritual connection. To the Mupuche in Chile, Weluwitraw (also known as Orion) represents a traditional Mapuche sport consisting of two men who have a rope in common tied to their necks pull in opposite directions until one of them manages to win, dragging the opponent. Orion's belt represents one of the men and the Orion Nebula represents the other, with the sword representing the rope between the two. In some sectors it is used as a temporal and spiritual variable. In other words, time of year observed and personal spiritual connection can change meaning.

Indigenous astronomy highlights the process of using long term (generational) observations of the sky to build knowledge within a cultural context and locational context. The cultural context is important because the observations of the sky and the implications were/are part of Indigenous life and livelihoods.

### V.     Why talk about Indigenous astronomy in a planetarium?

It is, first, important to acknowledge that narratives, language, knowledge and culture are essential in astronomy communication in planetariums. Currently, there is growing hunger from communities to learn more of Indigenous Astronomies (P. Harris and Matamua 2012) with Indigenous narratives being fundamental to understand this codified Indigenous knowledge (Hikuroa 2017). However, by uncritically centering Greek and Arabic constellations, star names and narratives, we are locating the night sky in Greek and Arabic narratives, language, knowledge and culture. This intellectually positions the planetarium in Greek and Arabic lands, rarely the lands of the planetarium. At other latitudes, constellations appear rotated or even upside down. In this way, planetariums can further disconnect the audience from their night sky. This section explores this sense of distance between the audience and the night sky created by Western-Eurocentric Astronomy.

When located in other lands, such as Aotearoa New Zealand, the irrelevance of the Western-Eurocentric narratives in astronomy becomes even more apparent. As an example, there are no scorpions in the Pacific Islands, therefore Scorpius, the scorpion, has no relevance to the lives of indigenous peoples of the Pacific. Contrarily, this constellation is seen as Te Mataunui-a-Maui, Maui's fishhook. A fishhook is a familiar shape to the seafaring peoples of the islands in the Pacific, and recognizable as such in their night skies. Further, the heliacal rising of Rehua (Antares), the brightest star in this constellation (Scorpio/Mataunui-aMaui), occurs in mid-December in Aotearoa New Zealand. Rehua rising signals the height of summer and ripens the berries in the trees within Māori narratives. Again, the indigenous astronomy is culturally embedded as well as situated within the local context. A wealth of relevant star knowledge can be observed connected to the environment and reinforces our relationship with the stars in the night sky.

Many astronomers and astrophysicists recall a visit to a planetarium being the start of a life-long relationship with the night sky sparking curiosity, excitement and careers. This speaks to the main purpose of planetariums - sparking a relationship with the night sky by representing the night sky, first and foremost, where you are. This is especially important in towns and cities where light pollution renders a star-filled sky invisible. The planetarium projects on to the dome what is out there in the sky, unseen and disconnected, communicating excitement and connection. If the intention of planetariums is to spark/build relationships with the night sky, closing the palpable distance between us and the stars, then locating astronomy in place through Indigenous astronomy strengthens the relationship between the audience and the night sky, and closes that sense of distance.

One of the key distinctions between Western astronomy and Indigenous astronomy as an Indigenous Knowledge System, is the acknowledgement of the relationality between the stars in the night sky and the rest of the environment. Indigenous star knowledge often reflects seasonality specific to place (Clarke and Harris 2018). The heliacal rising of particular stars can indicate months and the appearance of constellations at specific positions in the sky can indicate seasons (Matamua 2017) and reinforce our connection to the stars through our observation of this environmental seasonality.

The seasonality embedded in Western-Eurocentric narratives are out of sync with most local environments. As mentioned above in Aotearoa New Zealand, Rehua (Antares) returns to the night sky in December near midsummer signaling the driest hottest time of year, this time is dominated by the Western association with the constellation Aquarius reigning from mid-January to mid-February. Both occur in the height of summer, when there is rarely rain. In contrast, the Western narrative of Aquarius 'The Water Bearer' is seasonally distant and requires the audience to mentally locate themselves away from the local night sky and planetarium.

Although this section focuses on the obvious cognitive dissonance between Northern and Southern Hemisphere astronomies, with a particular focus on Aotearoa New Zealand, similar examples could be used from across the globe, including the Northern Hemisphere. In some locations the differences can be subtle. The differences can also be as diametrically opposed as rain and drought. Including Indigenous astronomy in planetarium presentations, for Indigenous and non-Indigenous peoples, achieves the objectives of planetariums in a unique way specific to place. Indigenous astronomy can strengthen the relationship of the audience with the night sky, with the local environment and, most profoundly, with the local Indigenous peoples.

### VI    Bringing Indigenous Astronomy into Planetarium Programs

Planetarium programs such as Stellarium now feature a large number of culturally-specific packages under "Starlore". These default plug-ins feature star names, constellation art, and cultural traditions for dozens of Indigenous communities and ancient cultures. Examples include Maor, Lokono, Maya, Tongan, Navajo, Mongolian, Tupi, Ojibwe, D/Lakota, Boorong, Aztec, Egyptian, Norse, and Chinese (among many more). Additional plug-ins can be developed through Stellarium, enabling the inclusion of Indigenous content from anywhere in the world.

Planetaria should include Indigenous content, or develop Indigenous-specific programs, whenever possible. Accomplishing this means adhering to established guidelines rather than the development of ad-hoc programs without relevant community consultation. Those guidelines are as follows:

1. Ensure all planetarium staff are properly trained in cultural competence (ACECQA, 2014);
2. Identify local communities/tribes with astronomical knowledge that could be included in planetarium programs or displays;
3. Engage in due diligence to see what knowledge is available, understanding restrictions, and identifying the key elders or representative organizations;
4. Follow established protocols for approaching and working with those communities/tribes;
5. Ensure constant consultation with the relevant community/tribe who must give final approval for all content and delivery options;
6. Ensure mutual benefits are in place for that community/tribe. This may include payments to elders, profit sharing, education and outreach programs for the local community/tribe, employment opportunities, educational materials, etc.;
7. Ensure Indigenous voices are centered and provide programs for local Indigenous people to deliver content whenever possible;
8. Produce a signed MoU that sets out rules and guidelines for permissions, program content and delivery, and future alterations.

Some of these guidelines need to be unpacked for further clarification and illustrative examples from various parts of the world are used to show these protocols in action and their benefits. Educators need to be knowledgeable about local communities and be culturally competent so as not to promote stereotypes or false information, regardless of intent. These are among the most problematic and common issues that face educators that can have long lasting repercussions. Many regions have established protocols for working with Indigenous people (ATSIEB 2015). In Australia, this is set out by organizations such as AIATSIS - the Australian Institute for Aboriginal and Torres Strait Islander Studies (AIATSIS 2015).

Educators should work with local communities on which the planetarium sits. While it is useful to discuss Indigenous astronomy across a specific country or region, it is critically important to include local knowledges and voices. Indigenous astronomy knowledge has to be collaboratively developed with tribes so accurate information can be shared. This collaboration must ensure that final approval for the delivery of all content is approved by the representative Indigenous body. Some communities have suffered greatly from colonization and much of their knowledge may be fragmented or lost, so this careful process is in response to this. Mapuche star knowledge, for example, was passed from generation to generation through oral tradition. There are no reliable records of Mapuche writing that reveals pre-colonization knowledge. In most cases Indigenous Astronomy is held as a sacred narrative that acknowledges a natural cosmic order that in turn determines a unique way of life, unique to the people involved, and in some cases this information may be sensitive or cannot be shared with non-tribal members. This may not always be possible, but it should be the desired focus. Tribal protocol contains appropriate tribal and cultural integrity and restrictions should be acknowledged and followed, parallel to the CARE principles for Indigenous data governance (Carrol et al. 2019). A clear MoU (Memorandum of Understanding) should be developed between the planetarium and the collaborating community/tribal organization. This should lay out the protocols for knowledge sharing, use of language and terminology, consideration of restrictions, protocols of future amendment, recognition of knowledge holders, plans for centering Indigenous voices, and the mutual benefits for the community.

The protocols and practices for working with Indigenous people are regionally specific. Many Aboriginal communities restrict certain knowledge to men, women, and/or senior initiated elders. These rules may vary from culture to culture. For example, the celestial emu is a motif featured in the traditions of a majority of the 350+ language groups spread across the Australian continent (e.g. Fuller et al. 2014a). Knowledge about the celestial emu may be public with some communities but deeply secret with others. Some elements may be restricted to men, while others are restricted to women (Michaels 1985). Some elements may be restricted to senior initiated elders, in sum certain indigenous knowledge may be freely shared or not dependent upon the ethnic group and their rules governing such knowledge.

There may also be restrictions on showing images and names of people who have passed away (NSLHD, 2015). In general, showing images and speaking names of people who have passed away is considered taboo in many Aboriginal communities across Australia, so this must be considered when working with communities/tribes on programs and content delivery. This also means protocols need to be established to accommodate potential alterations in the content delivery in the future.

As opposed to the practice of archeoastronomy, which tends to focus on cultures in the ancient past (which may no longer exist, or currently exists in a much different form), Indigenous Astronomy focuses on contemporary, living people and cultures (Medupe 2015). Therefore, it is important to acknowledge and understand the importance of focusing on Indigenous Knowledge as a living entity, to focus on the importance of Indigenous language, and to center Indigenous voices whenever possible. It is also critical that any program be developed and delivered in such a way as to not be derogatory towards Indigenous astronomy through comparison of Western astronomy (Ruggles 2010). When comparison is presented the equivalence of Indigenous astronomy to non-instrument Western astronomy should be emphasized. Indeed, when all the instruments are put aside, what shines bright is simply relationship to sky.

For the Sir Thomas Brisbane Planetarium in Brisbane, Australia, a large permanent wall-display was developed to feature Aboriginal and Torres Strait Islander astronomy, with a focus on the scientific elements of this knowledge (GC2018CGC 2018). Entitled *STARLORE*, it features examples from three different communities, showing diversity in terms of geography and representation: Wardaman (Aboriginal, Northern Territory), Euahlayi (Aboriginal, New South Wales), and Meriam (eastern Torres Strait, Queensland). It was decided not to show photographs of each elder, but elders gave their permission for their names to be shown even after death. Each section features a small map of Australia that shows the location of each community. The elders spoke at the launch and were centered in the media regarding the display. A similar display was developed at Perth Observatory (2018) in Western Australia entitled *Worl Wangkiny*, led by Aboriginal elder Dr Noel Nannup.

In the case of STARLORE, these three communities were selected because senior elders in those communities had published a significant amount of their astronomical knowledge (e.g. Cairns and Harney 2003; Fuller et al. 2014a,b,c; Guedes et al., 2018; Hamacher et al. 2018, 2019) and Senior Elders or Boards of Elders in those communities were able to give permission for that Knowledge to be shared and displayed. Given the relatively small size of those communities, such a thing was possible. With very large communities, there may not exist a single central body that can give blanket permission for knowledge to be shared. In select cases, any shared knowledge may need to focus on smaller sub-groups within the larger community, such as clan, family, or dialect groups.

Protocols vary from tribe to tribe. These protocols include and identify specific times when it is appropriate to tell stories of the sky. For the Dine, this is closely tied to the natural order of lunar, stellar and seasonal cycles, usually spanning the winter months from late September to mid-March. There are also restrictions on viewing celestial events such as lunar and solar eclipses that vary among and within tribes. When Planetarium educators show programs that are time specific, they should definitely acknowledge and follow the Indigenous Astronomer guidelines.

Restrictions will become clear as one works with different tribes. These are very important to the sharing of Indigenous astronomical information. In addition, it is always important to acknowledge the source of information (whether it is song or story) as coming from credible knowledge holders. For example, there is protocol around when star knowledge stories should be told. In Ojibwe one restriction is that certain stories should be told only when there is snow on the ground. As explained by Ojibwe elder W. Wilson, "Biboonkeonini – Wintermaker is a spirit that makes winter. Each season has certain spirits that make the season happen. Winter-only stories are told in wintertime because a person knows the Winter Spirit is there. No winter stories are told after the frogs wake up," (Lee et al. 2014: 27).

Other tribes, such as the Dine in Arizona, Utah and New Mexico, also have strict protocols involving when stories of the sky can be told (Maryboy and Begay 2017). For example, Winter Stories are closely linked to cosmic cycles of the Sun and Moon. Generally speaking, one only tells these stories from late September to mid-March. When the First Thunder of spring is heard, it is announced all over Navajoland, by radio or newspaper or word of mouth. That is the time plants awaken and animals come out of hibernation, having been stirred by the energies of the Thunder and other signs. It is the time at which Winter Stories can no longer be told (Maryboy and Begay 2005). This protocol is extremely important to be followed if a Planetarium plans to show Navajo stories. There is one time around the summer solstice when some of these stories may be shared, but for educational purposes only. If the protocol is not followed, a planetarium can be severely criticized by local Navajos. Information may also be restricted by gender, phenomena (e.g. taboos around viewing eclipses), time of day, location, or other factors that must be taken into consideration through the MoU.

Finally, it is critical that Indigenous people are able to speak for and about Indigenous Knowledge (Carnes 2011). The collaborating community must give clear permission about how their knowledge is presented, especially if the planetarium educators are non-Indigenous or not from the community sharing that knowledge. Examples of centering Indigenous voices may include naming specific elders who shared knowledge, recording their voices or videos so they can deliver it remotely, or nominating an appropriate person to deliver content if an Indigenous staff member is unavailable or non-existent.

If the planetarium has no Indigenous presenters or educators, significant efforts should be made to correct this. Programs have been developed at astronomy-related education and outreach facilities around the world that provide pathways for Indigenous people to be hired and trained as astronomers, educators, and science communicators. Sydney Observatory developed an Indigenous program (Wyatt et al. 2014) that brought in Aboriginal guides to deliver Indigenous programs. The guides delivered programs that were not only culturally and ethically appropriate, but they were able to draw from their lived experiences, which non-Indigenous educators cannot do in this context. If possible, the planetarium should provide programs that can be delivered in the relevant Indigenous language, enabling Indigenous educators to deliver programs to their communities in their language(s).

## VII    Conclusion

In 1999 the United Nations Educational, Scientific, and Cultural Organization (UNESCO) in conjunction with the International Council for Science (ICSU) held a World Conference on Science. The resulting report advocated global governments to support and promote understanding of traditional knowledge systems. First outlined in the Preamble, "All cultures can contribute with valuable scientific knowledge" (UNESCO 2003, 9), with greater detail the report states:

> *...traditional and local knowledge systems, as dynamic expressions of perceiving and understanding the world, can make, and historically have made, a valuable contribution to science and technology, and that there is a need to preserve, protect, research and promote this cultural heritage and empirical knowledge.*
> (UNESCO 2003, 14)

The report goes on to urge the scientific community to support, create dialogue, and build relationships, with "traditional societies and philosophers from all countries" (UNESCO 2003). Specifically:

> *Modern science does not constitute the only form of knowledge, and closer links need to be established between this and other forms, systems and approaches to knowledge for their mutual enrichment and benefit. A constructive intercultural debate is in order to help find ways of better linking modern science to the broader knowledge heritage of humankind.* (UNESCO 2003, 40)

The recommendation is for scientists to respect, sustain, and enhance traditional knowledge systems and that traditional knowledge should be integrated into interdisciplinary projects. Clearly the world's leading voice in science, education, and culture, UNESCO, understands the importance of widening the lens of science.

Lastly, an important concept guides this work in Indigenous Astronomy and museum collaboration, '*Etuaptmumk*' or Two-eyed seeing. Here are the words of two Mi'kmaw elders:

> *Two-Eyed Seeing is learning to see from one eye with the strengths of Indigenous knowledges and ways of knowing, and from the other eye with the strengths of Western knowledges and ways of knowing, and to use both these eyes for the benefit of all.* (Bartlett, Marshall, and Marshall 2012, 336)

In conclusion, it is important to remember that planetariums can provide immersive experiences where visitors can understand and connect to different cultures through the stars. Further, these experiences can be used to promote awareness that these stories are not just myths as they are often portrayed. These stories are a rich source of information about how ancient cultures lived and what was relevant and important to them. The stories are a rich source of scientific data that have been preserved in the stars and passed down from generation to generation orally. And in some cases the connection to the stars is a kind of a 'spiritual lifeline' or guidebook for the people to know: where they came from, what they are doing here, and where they are going. In Lakota, the stars were and are called "woniya of Wakaŋ Taŋka, the holy breath of the Great Spirit…when Lakota observed the movement of the Sun through their constellation, they were receiving spiritual instruction (Goodman 1992, 1)". In either case, the teaching that the stars are 'our oldest living relatives' cannot be understated. As illustrated in this quote from a tribal member after first learning about the Ojibwe constellations:

> *I used to look up and see the Greek constellations, like the Big Dipper or Leo the lion… but now I know that there are stars up there that are ours. It does something to me inside, to have that relationship with the stars. It's like finding a long-lost relative.* (Tibbetts 2010)

This document serves to support collaborative efforts by museums and planetariums that desire to integrate Indigenous astronomy and science content "in a good way" into their programming, content, and institution. The roadmap starts with tangible efforts to build authentic relationships with Indigenous knowledge keepers and Indigenous scholars. More than a side-note, the Indigenous voice should be allowed to work collaboratively with museum/planetarium staff to lead the Indigenous content. Ideally, museums would aim to work away from the 'add on consultant' model towards the more sustainable model of hiring full time positions such as 'Curator of Indigenous Scientific Knowledge Systems' or various staff positions. Ultimately, the aim of this work is increasing STEM/Informal Science Learning opportunities for Indigenous youth, adults, and communities, but also has the foundational goals of increased cultural pride, engagement in science, and community wellness. For the non-Native audience, there is great value in learning and practicing cultural agility. Standing together we have enormous reach and capacity.

Lee, Annette S., William Wilson, Jeffrey Tibbetts, and Carl Gawboy. 2014. *Ojibwe Star Map Constellation Guidebook: An Introduction to Ojibwe Star Knowledge*. Minneapolis, Minnesota: Native Skywatchers Press.

Little Bear, Leroy. 2000. "Jagged Worldviews Colliding." *Reclaiming Indigenous Voice and Vision*, Vancouver, , 77–85.

Maryboy, Nancy, and David Begay. *Sharing the Skies: Navajo Astronomy - Planetarium Full Dome Show*. Rice University, 2017.

Maryboy, Nancy, and David Begay. 2020.

Matamua, Rangi. 2017. *Matariki : The Star of the Year*. Huia Publishers.

Mawere, Munyaradzi. 2015. "Indigenous Knowledge and Public Education in Sub-Saharan Africa." *Africa Spectrum* 50 (2): 57–71.

Medupe, T.R., 2015. *Indigenous Astronomy in Southern Africa.* Handbook of Archaeoastronomy and Ethnoastronomy. Springer Science+Business Media, New York, pp. 1031-????.

Michaels, E. 1985. Constraints on Knowledge in an Economy of Oral Information. *Current Anthropology*, Vol. 26(4), pp. 505-510.

NSLHD, 2015. *Death and Dying in Aboriginal and Torres Strait Islander Culture (Sorry Business): A Framework for Supporting Aboriginal and Torres Strait Islander Peoples Through Sad News and Sorry Business*. Northern Sydney Local Health District, Sydney. URL: https://www.nslhd.health.nsw.gov.au/Services/Directory/Documents/Death%20and%20Dying%20in%20Aboriginal%20and%20Torres%20Strait%20Islander%20Culture_Sorry%20Business.pdf

Ogunniyi, Mb. 2005. "The Challenge of Preparing and Equipping Science Teachers in Higher Education to Integrate Scientific and Indigenous Knowledge Systems for Learners." *South African Journal of Higher Education* 18 (3): 289–304. https://doi.org/10.4314/sajhe.v18i3.25498.

Oxford English Dictionary. 2019. "Physics, n." In *OED Online*. Oxford University Press. http://www.oed.com/view/Entry/143140.

Perth Observatory, 2018. *Worl Wangkiny – Aboriginal Astronomy.* Perth Observatory. URL: https://www.perthobservatory.com.au/worl-wangkiny-aboriginal-astronomy

Ruggles, C.L.N., 2010. *Indigenous Astronomies and Progress in Modern Astronomy*. In Norris, R.P. and Ruggles, C.L.N. (eds) *Accelerating the Rate of Astronomical Discovery*, Special Session 5, IAU General Assembly, August 11-14 2009, Rio de Janeiro, Brazil.

Spieth, Jakob. 1911. *Die Religion Der Eweer in Süd-Togo*. Leipzig: Dieterich. https://sammlungen.ub.uni-frankfurt.de/dsdk/content/search/9348041?query=Dzerets

Tibbetts, Jeffrey. 2010. participant of a Native Skywatchers workshop.

UNESCO. 2003. "Science for the Twenty-First Century-A New Commitment. Declaration on Science and the Use of Scientific Knowledge-Science Agenda-Framework for Action." World Conference on Science. Budapest Hungary, June 26-July 1, 1999. Paris: UNESCO.

Unsöld, A. and Baschek, B., 2001. *The New Cosmos: An Introduction to Astronomy and Astrophysics*. Springer-Verlag, Berlin Heidelberg

Wyatt, G.; Stevens, T.; and Hamacher, D.W., 2014. Dreamtime Astronomy: a new Indigenous program at Sydney Observatory. *Journal of Astronomical History and Heritage*, Vol. 17(2), pp. 195-204.